\documentclass[a4paper,12pt]{article}
\usepackage{graphics}
\usepackage{graphicx,axodraw}
\usepackage{amsmath,amsxtra,amssymb,latexsym, amscd,color}
\usepackage{color}
\usepackage{axodraw}
\begin{document}
\title{{\huge \textbf{ Was the Higgs boson discovered?}}}

\author{
{\bf Nguy$\tilde{\hat{\mbox{e}}}$n Anh K$\grave{\mbox{y}}$}$^{\dagger,\ddagger}$
and 
{\bf Nguy$\tilde{\hat{\mbox{e}}}$n Thi H$\grave{\hat{\mbox{o}}}$ng V$\hat{\mbox{a}}$n}$^{\dagger}$\\[3mm] 
$^\dagger$\textit{Institute of physics},\\ \textit{Vietnam academy of science and technology (VAST)},\\ \textit{10 Dao Tan, Ba Dinh, Hanoi.} \\[3mm] $^\ddagger$\textit{Laboratory of high energy physics and cosmology},\\ \textit{Faculty of physics}, \textit{College of science (VNU university of science)},\\ \textit{Vietnam national university},\\ \textit{334 Nguyen Trai, Thanh Xuan, Hanoi.}}


\maketitle

\begin{abstract}
The standard model has postulated the existence of a scalar boson, named the Higgs boson. This boson plays a central role in a symmetry breaking scheme called the Brout--Englert--Higgs mechanism (or the 
Brout--Englert--Higgs--Guralnik--Hagen--Kibble mechanism, for completeness) making the standard model realistic. However, until recently at least, the 50-year-long-sought Higgs boson had remained the only particle in the standard model not yet discovered experimentally. It is the last but very important missing ingredient of the standard model. Therefore, searching for the Higgs boson is a crucial task and an important mission of particle physics. For this purpose, many theoretical works have been done and different experiments have been organized. It may be said in particular that to search for the Higgs boson has been one of the ultimate goals of building and running the LHC, the world's largest and most powerful  particle accelerator, at CERN, which is a great combination of science and technology. Recently, in the summer of 2012, ATLAS and CMS, the two biggest and general-purpose LHC collaborations, announced the discovery of a new boson with a mass around 125 GeV. Since then, for over two years,  ATLAS, CMS and other collaborations have carried out intensive investigations on the newly discovered boson to confirm that this new boson is really the Higgs boson (of the standard model). It is a triumph of science and technology and international cooperation. Here, we will review the main results of these investigations following a brief  introduction to the Higgs boson within the theoretical framework of the standard model and Brout--Englert--Higgs mechanism as well as a theoretical and experimental background of its  search. This paper may attract interest of not only particle physicists but also a broader audience.  
\end{abstract} 
\maketitle
 
\section{Introduction}
~

\begin{figure}
\begin{center}
\includegraphics[width=.7\textwidth, angle = 0]{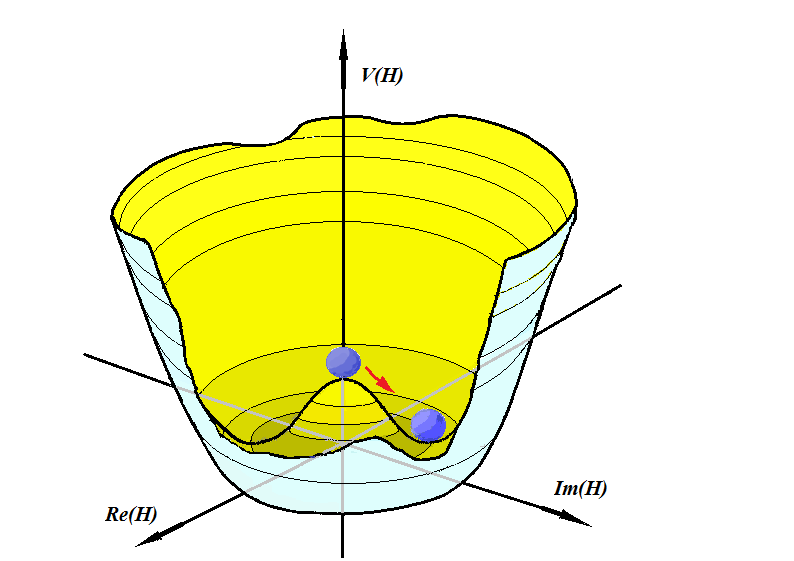}
 \caption{Spontaneous breaking of $U(1)$ symmetry when the Higgs potential has no minimum but unstable local maximum at the origin $\langle|H|\rangle =0$.}
 \label{ssb}
 \end{center}
\end{figure}

The standard model (SM) is a model of elementary particles and their interactions (for the basics of the SM, see, for example, \cite{Griffiths:1987tj,Cheng:1985bj,Peskin:1995evj,Quang:1998yw}). The SM combines (not ``unifies" yet) the strong interaction with the electro-weak (EW) interaction which unifies the weak interaction and the electromagnetic (EM) interaction (the latter in turn unifies, as well known, the electric- and the magnetic interaction). It has proven to be an excellent model as it can explain many phenomena and many of its predictions have been confirmed by the experiment. 
The SM is a model based on the gauge symmetry principle according to which all particles in the model would be massless (at the beginning at least) as the presence of a massive particle would violate the gauge symmetry. However, in reality not all elementary particles are massless, thus, the SM cannot be a realistic model if its gauge symmetry is preserved unbroken. In other words, the gauge symmetry of the SM must be somehow broken, unless there is another way to circumvent this problem. Therefore, one must find a mechanism to break the starting symmetry (down to a necessary smaller symmetry) and generate particles' masses. Fortunately, one such mechanism, called the  Brout--Englert--Higgs--Guralnik--Hagen--Kibble mechanism, or, just the Brout--Englert--Higgs (BEH) mechanism, for short, was indeed suggested fifty years ago, in 1964, by three independent groups: R. Brout and F. Englert (from Belgium) \cite{Englert:1964et}, P. Higgs  (from Scotland) \cite{Higgs:1964pj}, and G. Guralnik, C. Hagen and T. Kibble (from England and USA) \cite{Guralnik:1964eu}, following an earlier (non-relativistic) version suggested by P. Anderson in 1962 \cite{Anderson:1963pc}. According to this mechanism, particles adopt masses after a spontaneous symmetry breaking (SSB) reducing the original, bigger, symmetry emerging in a higher energy state to a smaller symmetry more stable, or even unbreakable, at a lower energy level. In general, a SSB can occur with a physical system when the symmetry of the lowest state (vacuum) is smaller than that of the equations (describing the underlying laws) of the system. A symmetry can be continuous (i.e., dependent on continuous parameters) or discrete.  Figure  \ref{ssb} gives an illustration of a spontaneous breaking (SB) of a continuous symmetry which is the case considered here. A symmetry can be global (independent of coordinates) or local. A SB of a global symmetry was considered in 1960 by Y. Nambu \cite{Nambu:1960xd}, however, it is accompanied by massless fictive particles called Nambu-Goldstone bosons (NGB's) \cite{Nambu:1960xd,Goldstone:1961eq}. This problem is solved in the case of SB of local symmetry (gauge symmetry) when NGB's can be absorbed by some gauge bosons to make the latter massive. Treated as a gauge theory, a fundamental interaction is carried by the so-called gauge bosons which are massless in case of an unbroken gauge symmetry and massive in case of a broken gauge symmetry. Long-range interactions are carried by massless gauge bosons but massive gauge bosons can carry only short-range interactions (the inverse assertion is not always correct). The strong interaction is a special case as its carriers, the gluons, are massless but it is a short-range interaction due to the confinement phenomenon. In general, the range of an interaction is proportional to the inverse mass of its carrier, for the strong interaction between nucleons the r$\hat{\mbox{o}}$le of this mass is played by the mass of a (neutral) $\pi$ meson considered as a carrier of an effective interaction residual from a more fundamental and much stronger interaction mediated by gluons between quarks. In a small (zero) mass limit the strong interaction has a chiral symmetry the spontaneous breaking (due to a quark condensation) of which generates almost the whole mass of a baryon composed of much lighter quarks.\\

The SSB phenomena are very common as they can also occur in different areas beyond particle physics, such as cosmology (closely related to particle physics), ferromagnetism, superfluidity, superconductivity, Bose-Einstein condensation, etc. In the Universe evolution, a SSB may occur when the temperature goes down \cite{Linde:2005ht}. A transition, for example, shortly after the Big Bang, from a hot, more symmetric, phase to a cooler, less symmetric, phase, can be treated as a kind of SSB. The maximal symmetry (not observed today but assumed to be present at a sufficiently high temperature in an early stage of the Universe) is broken down to a smaller and smaller symmetry when the Universe gets cooler and cooler. It is worth noting that this symmetry breaking process may not happen gradually or continuously but at definite ranges of temperature (energy). 
A (global) SSB could be observed in a ferromagnetic phase transition being a spontaneous magnetization when temperature decreases and passes a critical value $T_c$ (Curie temperature, in this case) \cite{ryder:1996}. For a three-dimensional case, this process spontaneously breaks the global symmetry $SO(3)$ to $U(1)$. Here, the NGB's are magnons, which are quanta of the spin waves. In  superconductivity \cite{Bardeen:1957kj,Bardeen:1957mv}, a SSB is caused (the symmetry $U(1)$ is broken) by a condensation of Cooper pairs leading to a gap in the Fermi surface. This gives rise to a non-zero effective mass of photons, making the magnetic field short-ranged, and, thus, preventing the latter to penetrate into the superconductive medium (Meissner effect) \cite{Anderson:1963pc}. The problem of the photon effective mass was earlier discussed as well by V. Ginzburg and L. Landau in their macroscopic description of superconductivity as SB of the electromagnetic gauge symmetry \cite{Ginzburg:1950sr}. The illustration of SSB given in Fig. \ref{ssb} is for a global and Abelian $U(1)$ symmetry but it can give an image for the case of a local and non-Abelian symmetry as in the standard model. \\

Applied to Glashow's electroweak unified model \cite{Glashow:1961tr}, an ingredient of the later created SM, by S. Weinberg \cite{Weinberg:1967tq} and A. Salam \cite{Salam:1968rm}, the BEH mechanism has brought a great success to the SM in explaining the particle masses and other phenomena (at least until an energy of about 200 GeV). It was also Weinberg \cite{Weinberg:1967tq} who first showed that fermions, i.e., the matter, could get masses via the BEH mechanism when the Higgs boson acquires a non-zero vacuum expectation value (VEV). However, without further driving works by G. 'tHooft and M.  Veltman \cite{'tHooft:1971rn,'tHooft:1972fi} and an efficient ``propaganda" \cite{Lee:1972fj}--\cite{Fujikawa:1972fe} by B. Lee (who coined and popularized the name \textit{``Higgs boson"}), the SM would remain unrealistic and unattractive. 
Proving the renormalizability of a spontaneously-symmetry-broken gauge theory \cite{'tHooft:1971rn,'tHooft:1972fi}, 'tHooft and Veltman made Glashow-Weinberg-Salam's electroweak model calculable and testable, thus, more realistic and attractive. \\

 In the SM with the incorporation of the BEH mechanism, particles can acquire masses via their couplings to an additionally introduced scalar field called briefly the Higgs field inducing a spontaneous symmetry breaking due to a condensation of the Higgs field, or in a more specialized language, when the Higgs field develops a vacuum expectation value (VEV). With treating the Higgs field as a condensate filling the whole space, these couplings (interactions of the considered particles with the Higgs field) are like walking through a molasses medium where the motion becomes heavier and slower, that is, a particle moving feels more weight. In particular, an interaction in such a medium can become shorter-ranged as the speed of the interaction carriers, looking heavier, is smaller than that in an empty space. Since the weak interaction is short-ranged, its carriers (weak-gauge bosons) could be expected to be massive, while  the electromagnetic interaction is long-ranged, its carriers, the photons, $\gamma$, must be massless.\\

The SM Higgs field, or just the Higgs field for short, is a two-component complex field having four degrees of freedom (DOF's). Through a SSB from the EW symmetry $SU(2)_L\otimes U(1)_Y$ to the stable EM symmetry $U(1)_{EM}$, three of these four DOF's (corresponding to Nambu-Goldstone bosons \cite{Nambu:1960xd,Goldstone:1961eq} in the case of a global symmetry breaking) are ``eaten" by the weak-gauge bosons, denoted as $W^\pm$ and $Z$, to make the latter massive as required by the short-range nature of the weak interaction, while the photons, $\gamma$, the quanta of the EM field, remain massless as the EM interaction is a long-range interaction. The fact that these massive gauge bosons, as predicted by the theory, were really discovered at CERN over thirty years ago \cite{Arnison:1983rp} -- \cite{Bagnaia:1983zx}, gives a strong support to the SM and the BEH mechanism. For the fermions, it was S. Weinberg who first showed that they could get masses proportional to the VEV of the Higgs field and the strengths of their interactions (Yukawa couplings) \cite{Weinberg:1967tq}. The remaining DOF of the neutral component of the Higgs field if existing could be discovered through its quantum excitation named the Higgs boson. Not only the latter is the only fundamental scalar particle in the SM but also it is a particle of new type different from other SM particles which are of either matter type (such as electrons, neutrinos and quarks) or interaction-mediating type (such as photons and gluons)\footnote{Here, as mentioned, we do not consider physics beyond the SM, including models in a higher-dimensional space-time where the Higgs boson can be treated as a gauge boson component on an extra-dimension direction, although the latest LHC results may shed light on these called also Higgs-gauge unification models (see, for instance, \cite{Gogoladze:2013rqa,Aad:2012cy} and references therein).}.
 However, at least until recently, the Higgs boson, had been the last but special particle in the SM not found yet experimentally, or, in other words, it had been the last missing but important piece of the SM. Therefore, searching for the Higgs boson is a crucial task of the experimental particle physics as the existence or the non-existence of the Higgs boson could decide the fate of the SM and a realization of the BEH mechanism. In particular, searching for the Higgs boson is an important mission, even one of the main goals, of the LHC ({\it Large Hadron Collider}), the most expensive, most powerful and largest particle accelerator ever built.\\

The LHC has several detectors, among which the ATLAS ({\it A Toroidal LHC Apparatus}) detector and the CMS ({\it Compact Muon Solenoid}) detector are the two biggest and general-purpose ones. On 4 July 2012, the ATLAS and the CMS collaborations released a piece of breaking scientific news announcing the discovery of a new boson  of a mass of around 125 GeV hoped to be the Higgs boson of the standard model \cite{Aad:2012tfa,Chatrchyan:2012ufa}. As the Higgs boson is the last missing SM particle, if the newly discovered boson is identified with the Higgs boson, the SM becomes a model with all its particle content confirmed experimentally. After the discovery, to check if the new boson is really the Higgs boson, it has been intensively and extensively studied so that its properties can be precisely determined, and, thus, its nature can be exactly established. Here we will briefly review the process of searching for the Higgs boson and the investigation on the new boson. 
Since the first observations of the new boson in 2012 via its decays to gauge boson pairs ($\gamma\gamma$, $WW$ and $ZZ$), its mass, spin-parity ($J^P$) and other characteristics have been more precisely measured or determined by different collaborations: ATLAS, CMS and Tevatron. In investigating the new particle, besides determining its couplings to the gauge bosons via the above mentioned decays, one next important step which much be done is to see if it also couples to fermions and if these couplings fit the SM couplings between the Higgs boson and the fermions providing masses to the latter. Another problem is to determine if the new particle has the same spin-parity of the Higgs boson which is a scalar. According to the measurements, done by ATLAS, CMS and Tevatron for last over two years, these questions get a positive answer. \\

All the results obtained so far support the newly discovered particle to be the long-sought Higgs boson, thus, from now on, for further convenience, the new particle can be referred to as the Higgs boson. This particle has a mass of about 125 GeV, spin-parity $J^P=0^+$ and its couplings to gauge bosons and fermions are consistent with those of the Higgs boson in the SM. The discovery of the Higgs boson is important because not only it is crucial for the SM and the BEH mechanism providing masses to particles, thus, to different ingredients of the Universe and the latter as the whole, but also it shows that an elementary scalar particle exists in the Nature. The existence of the Higgs boson is very meaningful for particle physics and other fields 
as all other scalar particles found so far are not elementary but composite. This discovery may also have cosmological and 
other consequences (for example, there is a hypothesis in which the Higgs bosons are inflatons \cite{Guth:1980zm} although there are later arguments against it \cite{Guth:1997wk, Ijjas:2013vea}) which cannot be discussed within this concise review but may be found elsewhere (see, for example, \cite{Linde:2005ht, Jegerlehner:2014mua} and references therein). \\ 

In the framework of this review we are able to discuss only selected moments in introducing, searching for and investigating the Higgs boson. It is far from being complete to cover all aspects of such a widely and intensively investigated topic. The present review being an extended version of \cite{vanky:2014bmt} devoted mainly to the results obtained by ATLAS, also discuss results of other experiments and other matter beyond particle physics.\\          

The SM can explain many but not all things. The problems such as neutrino masses and oscillations, CP-violation, the number of generations, dark matter and dark energy, etc., which are beyond the  ability of the SM to solve, call for an extension of the latter. However, the limited length and scope of this review do not allow us to discuss physics beyond SM where additional scalar (Higgs-type) fields may be needed.\\

For a plan of this review, before going to more physical discussions in section 3, let us make in the next section a technical overview on the LHC to give a general idea on its structure and operation as well as how a particle can be detected and investigated.
%
%
\section{LHC in brief} 
\vspace*{2mm}
\subsection{General information of the LHC}
~

Let us first make a brief description of the LHC with the ATLAS detector described in more details as an example of one of the LHC detectors playing a main role in searching for and discovering the Higgs boson. Most information presented in this section can be also found in \cite{vanky:2014bmt}. More information about these facilities are given in the official websites of the LHC and the ATLAS collaboration \cite{lhc,atlas}.\\

The LHC, the biggest (in size and cost) and most powerful (in collision energy) particle accelerator ever built by human, is located at CERN ({\it European organization for nuclear research}), which is the world's leading laboratory for particle physics. It is installed in a 27 km long orbicular tunnel (the former LEP tunnel) at a depth of 50 -- 175 m under the French-Swiss border near Geneva. The LHC was designed to accelerate and collide two proton beams at a center-of-mass energy (CME) of 14 TeV ($\sqrt{s}$ = 14 TeV). It was also designed to collide two beams of heavy ions (\textit{Pb}) accelerated to an energy of 575 TeV per ion but here we will work on proton-proton (\textit{p-p}) collisions only.\\

Protons, as hydrogen atoms stripped from electrons by an appropriate electric field, preliminarily accelerated to the energy 750 keV, are first injected into a \textit{linear accelerator} (LINAC 2) where they are accelerated to 50 MeV. Then, they are consecutively accelerated by the \textit{Proton Synchrotron Booster} (PSB), the \textit{Proton Synchrotron} (PS) and the \textit{Super Proton Synchrotron} (SPS) to reach energies 1.4 GeV, 25 GeV and 450 GeV, respectively, before being finally injected into the LHC where they are accelerated to record high energies in different stages of the LHC operation.\\ 

In the LHC first run (2009 -- 2012), the highest \textit{p-p} collision energy reached was 8 TeV ($\sqrt{s}$ = 8 TeV), a  record energy, while the designed maximal energy (14 TeV) is expected to be reached during the second run starting in the first months of 2015. Besides the record particle collision energies, the construction and the operation of the LHC have accomplished a number of scientific- and technological achievements such as the discovery of new particles, precision measurements of the SM parameters, a superstrong magnetic field (8.4 T, that is about 200 000 times stronger than the Earth's magnetic field), the highest vacuum ($10^{-10}-10 ^{-11}$ mbar, that is in the order of the vacuum on the Moon surface), the lowest temperature (1.9 K, that is lower than the temperature in the outer interplanetary space, 2.7 K), the highest temperature (5.5 trillion degrees Celsius, that is near 350 000 times of the temperature in the center of the Sun), etc. \\

Being a marvel of science and technology, the LHC program, including its designing, construction and operation, has involved collaborations from more than 10000 specialists from over 100 countries. To emphasize that the LHC program has proven to be a very successful scientific program we should mention that the LHC collaborations have discovered so far three new particles, not counting other scientific achievements. After the discovery of the Higgs boson by ATLAS and CMS, recently, LHCb, another LHC collaboration, announced the discovery of two new baryons being three-quark resonances \cite{Aaij:2014yka} which, however, are not a subject of discussion here in this paper.\\

The LHC also contributes to the computational science and technology. Its computing network - the {\it Worldwide LHC Computing Grid} (WLCG), is the world's largest computing grid.
Connecting over 170 computing centers from 40 countries in the world it is a driving factor behind the EGI ({\it European Grid Infrastructure}) which is a multi-scientific service. Without the WLCG which can process a huge amount of data, doing research with the LHC, specially, searching for the Higgs boson, would be unfeasible. Annually, it can store, distribute and analyze about 30 petabytes (30 million Gygabytes) of data produced by the LHC \cite{wlcg,grid:2011}. 
\vspace*{4mm} 
\subsection{LHC detectors}
~

\begin{figure}
\begin{center}
\includegraphics[width=.9\textwidth, angle = 0]{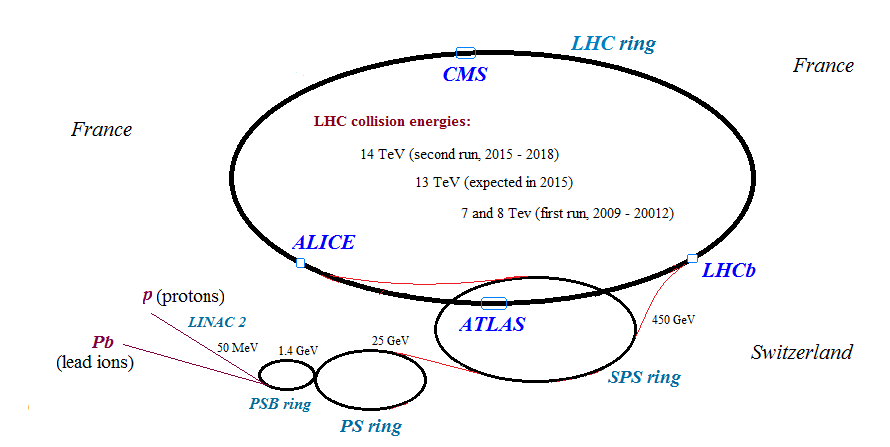}
 \caption{LHC ring and detectors.}
 \label{lhc-view}
 \end{center}
\end{figure}

The LHC has four main detectors, ATLAS, CMS, LHCb and ALICE (see Fig. \ref{lhc-view} for a general scheme of the LHC ring and the detectors), whereas ATLAS and CMS are the two biggest and general-purpose detectors  with the help of which the new boson was discovered. 
These detectors are huge and complex high-technological facilities based on the same operation principle and having similar general structures and purposes. They, when compared with each other, however, have some features. The ATLAS detector has four main sub-structures consisting in turn of many layers with a total mass of about 7000 tones and an overall size of about 25m (diameter) $\times$ 46m (length). The CMS detector is heavier (13000 tones) but smaller (15m $\times$ 22m), thus, the name {\it ``compact"}. Depending on construction materials, the ATLAS detector, compared with the CMS detector, has a more sensitive hadron calorimeter (thus, a better jet resolution)
but a less sensitive electromagnetic calorimeter (thus, a worse $e/\gamma$ resolution). The CMS inner detector surrounded by a 4T magnetic field has a better momentum resolution than the ATLAS inner detector surrounded by a 2T magnetic field but this design restricts the design of other components of the CMS. We have just briefly counted a few overall characteristics of the ATLAS- and the CMS detectors but the reader can consult \cite{Froidevaux:2006rg} to see more similarities and differences between these detectors.\\

The ATLAS and the CMS are very big collaborations with a wide research scope spreading from the test of the SM to searching for New Physics (physics beyond the SM): precision measurements of particle parameters and properties (compared with those predicted by the SM), search for the Higgs boson,  CP-violation (matter-antimatter asymmetry), extra dimensions, supersymmetry (boson-fermion symmetry), dark matter, etc. Each of these collaborations has attracted participation of more than 3000 scientists and engineers from more than 170 institutions in about 40 countries.\\

\begin{figure}
\begin{center}
\includegraphics[width=4.5in]{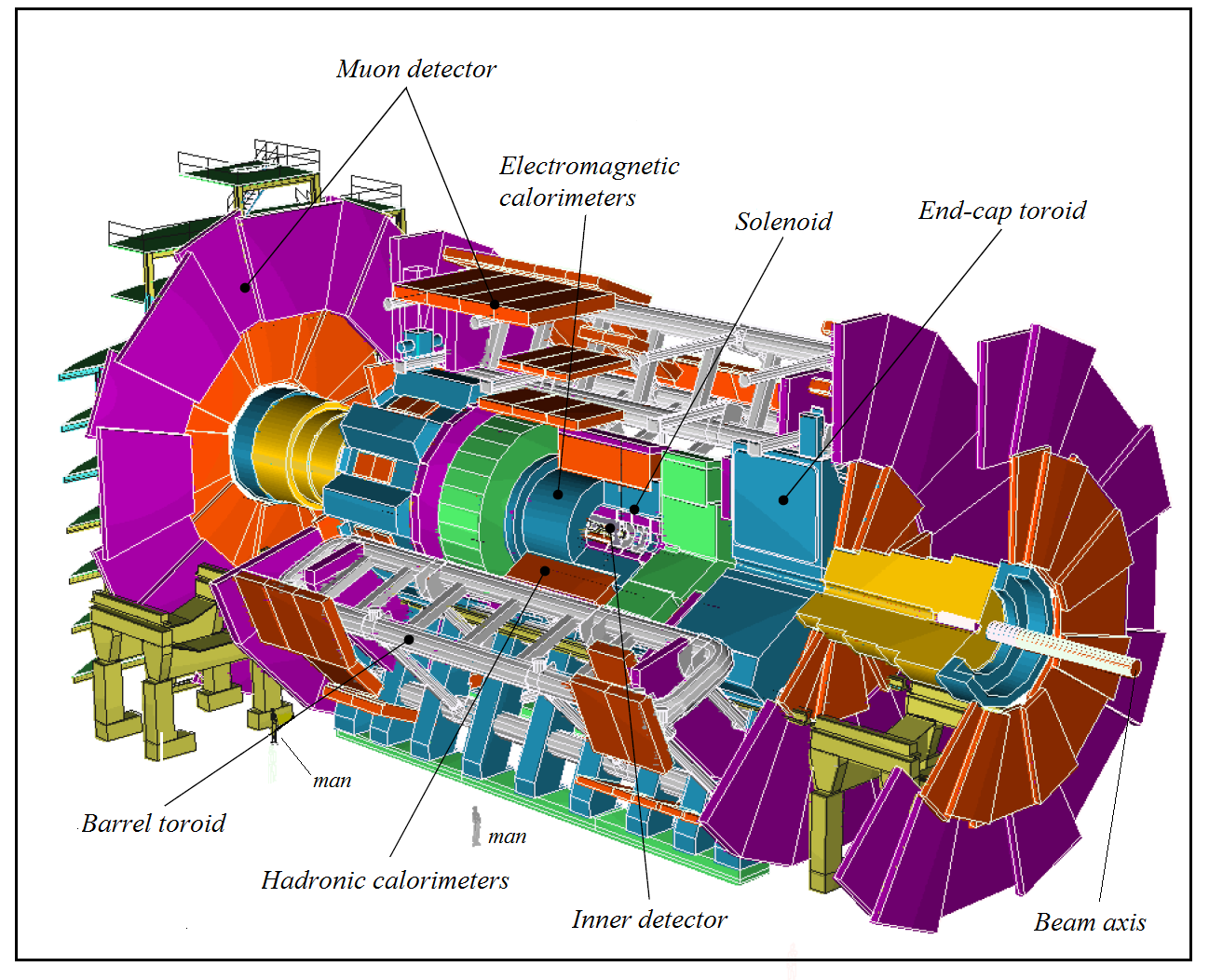}
\caption{The ATLAS detector layout \cite{atlas}.}
\label{atlas}
\end{center}
\end{figure}

As said above ATLAS and CMS are complex research facilities containing many components with different functions but here, for illustration, we will give a brief description of the ATLAS detector (see its layout in Fig. \ref{atlas}), as the CMS detector has similar general structure and basic operation principle, so that we can get a general idea how a particle can be detected and measured. The ATLAS detector has four main 
components representing ever-larger concentric cylinders, which, counted outward from the center, are the inner detector (ID), the calorimeters (CM's), the muon spectrometer (MSM) and the magnet systems (MS), all surrounding the proton beam axis at the center. Without going to details \cite{atlas,Froidevaux:2006rg}, let us recall general structures and functions of these components (presented also in \cite{vanky:2014bmt, van:2007ictp, van:2011cea}).\\

The basic function of the ID is to track and identify charged particles. It is the innermost component of the ATLAS detector surrounding the interacting point at the centre where collisions of proton beams take place and consisting of three high-resolution parts (the pixel detector, the semi-conductor tracker and the transition radiation tracker (TRT)), all surrounded by a solenoidal superconducting magnet system. The ID measures positions and momenta of charged particles in the pseudorapidity range $|\eta|<2.5$ (in which the TRT covers the range  $|\eta|<2.0$).\\

The next component of the ATLAS detector is the CM's surrounding the ID. Its function is to measure energies of (easily stopped) particles by absorbing these energies. This component of the ATLAS detector is composed of two sub-components: the  electromagnetic calorimeter (EC) and the hadronic calorimeter (HC). The EC is designed for high precision measurements of energies and locations 
(including trajectories) of particles sensitive to electromagnetic interaction such as photons and charged 
particles, while the HC measures the energies of those particles sensitive to the strong interaction such as hadrons. 
The HC has no high precision as the EC but it can measure the particles in the range $|\eta|<4.9$, which cannot be caught by the EC.\\

The outermost layer of the ATLAS detector is the MSM which is a very large system surrounding the CM's. The MSM has three parts: a set of large superconducting toroidal magnets, a set of chambers tracking with high spatial precision outgoing muons, and a set of chambers triggering particles with high time-resolution. This spectrometer is used to track 
outgoing muons being the only detectable particles which cannot be stopped by the CM's. It measures with a very high precision the paths and momenta of muons in the ranges $|\eta|<2.4$ (at triggering chambers) and $|\eta|<2.7$ (at tracking chambers).\\

The last component on our list is the MS placed in different places in the ATLAS detectors. These MS (solenoidal magnets and toroidal magnets) are designed to produce appropriate magnetic fields to bend trajectories of (charged) particles so that their momenta and charges can be determined. The solenoidal magnets, surrounding the ID, can produce 2 Tesla magnetic fields with a peak at 2.6 T, while the magnetic fields produced by the toroidal magnets 
around the MSM are 0.5 T (by the barrel coils) and 1 T (by the end-cap coils).\\ 

\begin{figure}
\begin{center}
\includegraphics[width=.8\textwidth, angle = 0]{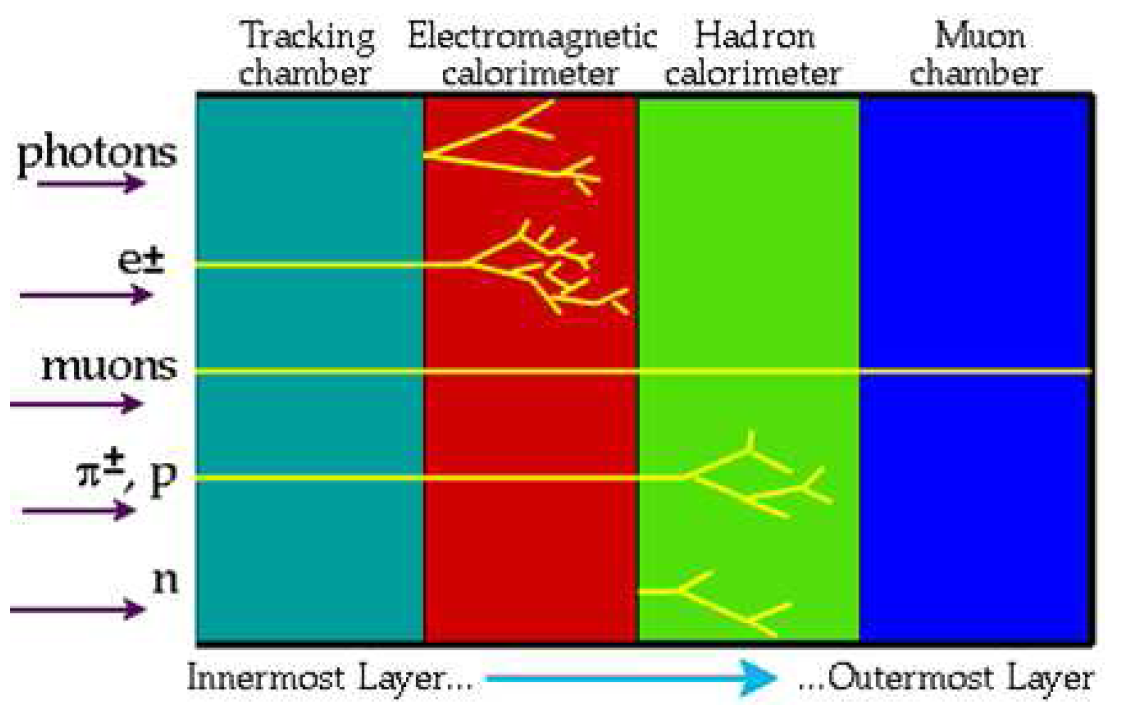}
 \caption{Simplified scheme of the ATLAS detector and particle detection  \cite{atlas}.}
 \label{traces}
 \end{center}
\end{figure}

For a summary and a general illustration of how a particle can be detected in the ATLAS detector (similarly, in the CMS detector), a simplified view on the structure of the ATLAS detector and particle detection is shown in Fig.  \ref{traces} (see more in \cite{lhc,atlas,van:2007ictp}). \\        

With its very large structure and high sensitivity, the ATLAS (along with the CMS) could detect for the first time a new boson which now, after a number of more precise investigations of its characteristics and properties (masses, spin-parity ($J^P$) and other production and decay information), can be almost confirmed to be the long-sought standard model Higgs boson. It is a scalar particle ($J^P=0^+$) having a mass of about 125 GeV, and coupled to gauge bosons and fermions (quarks and leptons) as expected for the Higgs boson. These investigations, summarized in the next section, have been done on the basis of the analysis of the data collected in 2011 (for $\sqrt{s}$ = 7 keV) and 2012 ($\sqrt{s}$ = 8 keV). It is expected the Run 2 of the LHC starting soon will give more results not excluding unexpected ones.

\section{Hunting and discovering the Higgs boson}
~

In this review we try to answer the question why and how the Higgs boson has been searched for and then the question of whether the Higgs boson was really discovered. As is well known, one of the central missions of the ATLAS- and the CMS collaborations is to search for the Higgs boson which plays a crucial r$\hat{\mbox{o}}$le in the symmetry breaking mechanism providing masses to particles. The fact that the gauge bosons ($W$ and $Z$), for example, get masses is very important as it makes the weak interaction short-ranged, otherwise, the structures like atoms, thus, the Universe, would not be formed, and many processes in the Nature such as the 
reactions in the Sun, making, in particular, the life on the Earth possible, would not occur. If the Higgs boson 
does not exist one must work out another symmetry breaking scheme or to deal with another mechanism to generate 
masses of particles (there have been such attempts but we do not discuss them here) because if all particles are massless our World would not exist at all or it would not exist as observed. The next argument making the Higgs boson important is that, if discovered, it would be the first real example of an existing fundamental scalar particle in the Nature and as discussed above it is believed to play an important role in particle physics and other branches of physics. Hence, the existence or the non-existence of the Higgs boson may decide the fate of the SM and other theories or models with the BEH mechanism incorporated in. All that explains why the Higgs boson has been one of the most sought after particles for nearly 50 years and its discovery could be classified to be among the most remarkable and important scientific discoveries in the last 100 years. Until the supposed discovery of the Higgs boson announced on 4 July 2012 the belief in its existence has increased over time as more and more predictions of the SM have been experimentally confirmed in its favour. This belief has created a strong motivation for searching for the Higgs boson and, hence, for building the LHC. \\

In order to identify a particle one must determine all its 
basic characteristics including its mass which may be in 
advance theoretically estimated or constrained by certain conditions. For the Higgs boson ($H$), until the discovery in 2012, different theoretical constraints and experimental results  (precision measurements of the SM model parameters) had  established bounds of its mass ($m_H$) which is one of the  fundamental parameters of the SM to be determined experimentally. \\

In the theoretical aspect, the Higgs boson mass, or simply, the Higgs mass, cannot be directly predicted by the SM but it can be constrained by, for example, the known SM parameters including masses of other particles such as the top quark and the gauge boson $W$ (or $Z$). The unitarity constraints \cite{Cornwall:1973tb,Cornwall:1974km,Llewellyn Smith:1973ey,Lee:1977eg} put an upper bound of the Higgs mass at around 1 TeV, while the validity of the SM up to the Planck scale is more rigorous requiring $m_H \leq $ 180 GeV (triviality bound) \cite{Ellis:2009tp}. When the stability of the Higgs potential is taken into account, the Higgs mass is also bounded from below at about 130 GeV (stability bound) \cite{Alekhin:2012py}. The lower bound may become smaller, at about 115 GeV \cite{Espinosa:1995se}, if a metastable electroweak vacuum 
is allowed. The fact that the Higgs mass 125 GeV is far from the triviality bound (that means there is no need of physics beyond SM until the Planck scale) but on the edge of the vacuum stability-instability, may lead to serious and interesting consequences.\\

In the experimental aspect, the Higgs mass can be determined or estimated indirectly by precision measurements \cite{Baak:2014ora}, or just ``measurements'', of electro-weak parameters such as Fermi constant, the top quark mass, the masses of gauge bosons \footnote{For example, see \cite{van:2011cea,Aad:2011fp} for one of the methods which can be used for precision measurement of the $W$ mass at the LHC.}, etc., or directly via a mass reconstruction from the Higgs decays. Different collaborations from LEP, Tevatron, LHC, etc. have made measurements to estimate the limits of the Higgs mass. They in general have been so far consistent with each other and with the SM. Let us briefly recall some results obtained before the discovery announced in 2012 (see, for example, \cite{Agashe:2014kda} for more precise information). The above mentioned measurements established or excluded ranges of a potential mass of a possible Higgs boson. By its shut-down in 2000 LEP (LEP-2) established a lower bound of the Higgs mass at about 114.4 GeV \cite{Barate:2003sz},
while the upper bound given by LEP, Tevatron and SLC was 152 GeV \cite{pmeasure:2012}. These bounds are quite consistent with the possible Higgs mass range 115 GeV $ <  m_H < $ 140 GeV derived by D$\emptyset$ and CDF collaborations at the Tevatron \cite{Group:2012zca}. The combined results \cite{Baak:2011ze} from LEP, D$\emptyset$ and CDF gave the Higgs mass range 115 GeV $ <  m_H < $ 135 GeV, slightly different from the range 115 GeV $ <  m_H < $ 130 GeV established later, but before 04 July 2012, by ATLAS and CMS \cite{atlas-seminar:2011, cms-seminar:2011}. The measurements put a possible Higgs mass in more and more narrow ranges until reaching the final value at about 125 GeV, a bit below the stability bound.\\

\begin{figure}
\begin{center}
\includegraphics[width=.8\textwidth, angle = 0]{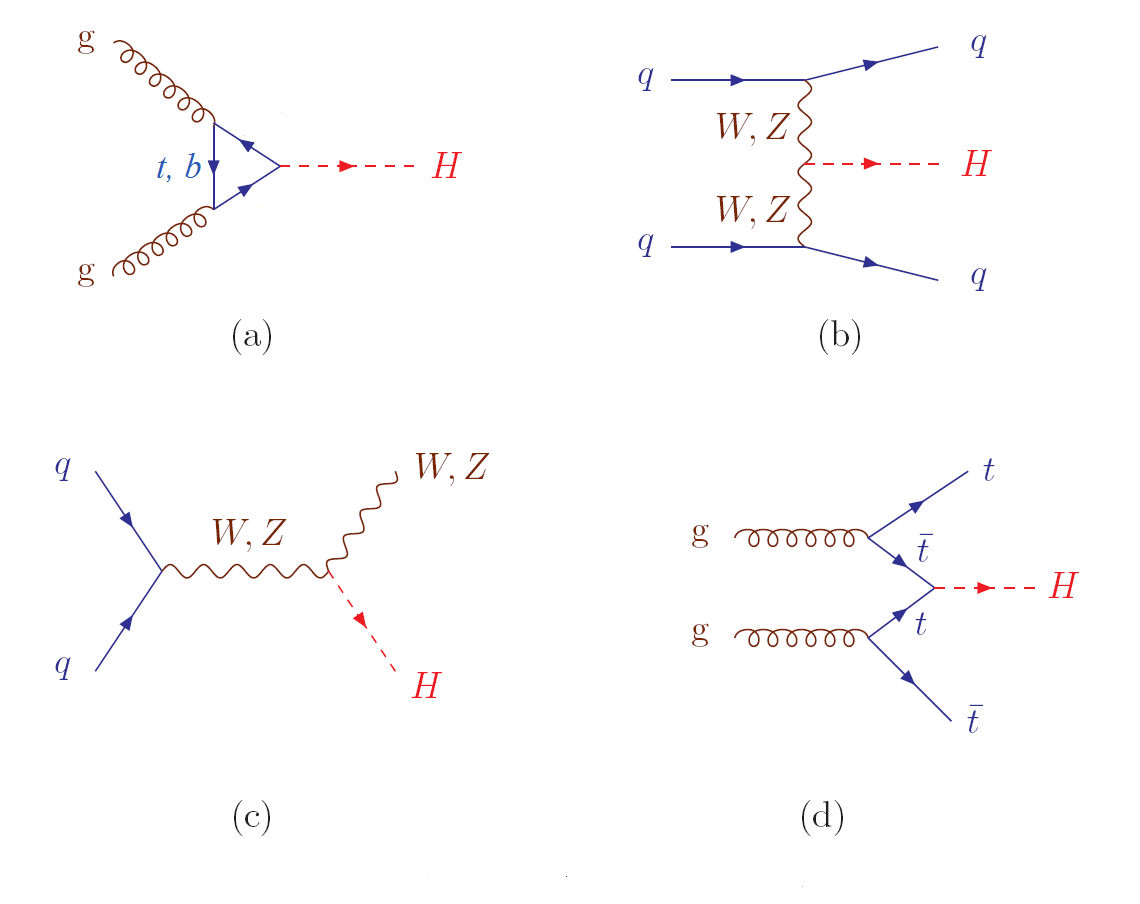}
 \caption{Higgs production diagrams: (a) {\it gluon fusion}, (b) {\it W/Z fusion}, (c) {\it Higgs strahlung}, (d) {\it top fusion}.}
 \label{Hprod}
 \end{center}
\end{figure}

At an accelerator like the LHC, the Higgs boson could be produced in various processes such as a gluon fusion, a weak-gauge-boson (W/Z) fusion and a Higgs production associated with a gauge boson ({\it Higgs strahlung}) or with top quarks (top fusion) \cite{Agashe:2014kda,Baglio:2010ae,Dittmaier:2011ti, Dittmaier:2012vm} 
(see Fig. \ref{Hprod} for corresponding Feynman diagrams). Among these processes, the gluon fusion 
followed by the gauge-boson fusion, is dominant. 
Once produced, the Higgs boson, because of its very short life time (of the order 10$^{-22} s $ for a mass around 125 GeV), decays 
immediately into lighter particles. According to the SM the possible decays of the Higgs boson could include 
$H\rightarrow \gamma \gamma$, 
$H\rightarrow  ZZ\rightarrow 4l$, 
$H\rightarrow WW \rightarrow l\nu l\nu$, 
$H\rightarrow \tau\tau$, 
$H\rightarrow bb$, 
etc., with branching ratios (BR's), relative uncertainties and mass resolutions given  in Table \ref{H-decays} (taken from \cite{vanky:2014bmt}) for the Higgs mass $m_H=125$ GeV  \cite{Agashe:2014kda,Baglio:2010ae,Dittmaier:2011ti, Dittmaier:2012vm}. A choice of an optimal channel
for the Higgs boson search depends on its sensitivity which in turn depends on several 
factors such as the cross section of the Higgs production, the branching ratio of the Higgs decay, the resolution 
of the reconstructed mass, the selection efficiency and the signal-to-background ratio ($S/B$). All these factors strongly
depend on a Higgs mass or a Higgs mass range. Thus, at a given Higgs mass or Higgs mass range, there is/are some channels more preferable than others for the Higgs search.\\
\begin{table}
\begin{center}
\begin{tabular}{||l||c|c|c||}
\hline
\hline
&&&\\
{\bf  {\boldmath $H$} decay channel} & {\bf Branching ratio} & {\bf Relative uncertainty}   & {\bf Mass resolution} 
\\ &&($\%$)&($\%$)\\ 
\hline
&&&\\
 $H\rightarrow \gamma\gamma$ & $2.28\times 10^{-3}$ &  {\large $^{ + 5.0}_{- 4.9}$}& 1-2
 \\ 
&&&\\
 \hline
&&&\\
 $H\rightarrow ZZ  ~(\rightarrow 4l)$ & $2.64\times 10^{-2}$ & {\large $^{+ 4.3}_{-4.1}$}& (1-2)
\\ 
&&&\\
 \hline
&&&\\
 $H\rightarrow WW ~(\rightarrow l\nu l\nu )$ & $2.15\times 10^{-1}$ & {\large $^{+ 4.3}_{-4.2}$} & (20)
\\ 
&&&\\
 \hline
&&&\\
 $H\rightarrow \tau\tau$ & $6.32\times 10^{-2}$ & {\large $^{+ 5.7}_{-5.7}$}& 15
 \\ 
&&&\\
 \hline
&&&\\
 $H\rightarrow bb$ & $5.77\times 10^{-1}$ & {\large $^{+ 3.2}_{-3.3}$}&10
\\ 
&&&\\
\hline
\hline
\end{tabular}
\vspace*{2mm}
\end{center}
\caption{Sensitive Higgs decay channels at the LHC for $m_H=125$ GeV \cite{vanky:2014bmt,Agashe:2014kda}.} 
\label{H-decays}
\end{table}

The Higgs mass measurements by ATLAS and CMS 
have been mainly based on 
the invariant mass reconstruction from the decay channels $H\rightarrow \gamma \gamma$ and $H\rightarrow ZZ
\rightarrow 4l$, where $l=e$ or $\mu$. 
As seen in Table \ref{H-decays}, the channels $H\rightarrow \gamma \gamma$ and $H\rightarrow ZZ  \rightarrow 4l$ have no large cross sections but they are preferred as ``golden" channels thanks to high mass resolutions 
(1-2$\%$) and clean signals. 
The other three channels are not excluded from use but there are  several difficulties. Compared with other channels, the channel 
$H\rightarrow WW \rightarrow l\nu l\nu$ has a large branching ratio but the Higgs mass 
resolution is very low (20$\%$) because of neutrinos produced in the final states (see Table \ref{H-decays}). The channels 
$H\rightarrow \tau\tau$ and $H\rightarrow bb$ have no clean signals because of a low mass resolution (15$\%$ and 10$\%$, resp.) and large backgrounds. The investigation of the latter channels, however, is important for determining whether the couplings of the Higgs boson (the new boson) to fermions are compatible with the SM (see below).\\

\begin{figure}[htbp]
\begin{center}
\includegraphics[width=.6\textwidth, angle = 0]{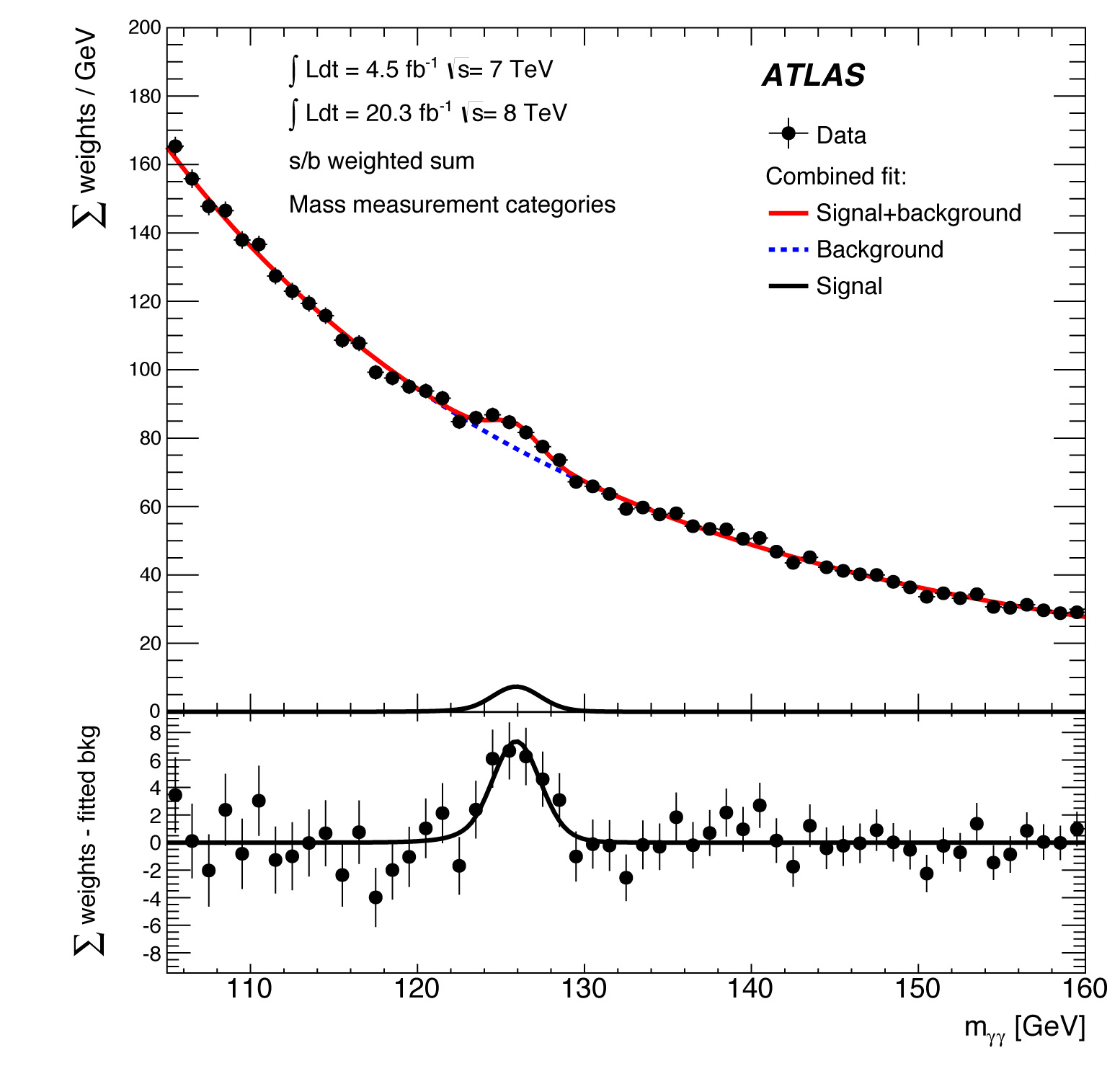}
\vspace{-2mm}
\caption{An invariant mass ($m_{\gamma\gamma}$) spectrum in decay $H\rightarrow \gamma\gamma$ for the combined 
$\sqrt{s} = 7$ TeV and 
$\sqrt{s} = 8$ TeV data and the mass range 105 -- 160 GeV \cite{Aad:2014aba,Aad:2014eha}. 
\label{mgamma}
\vspace{5mm}}
\includegraphics[width=.6\textwidth, angle = 0]{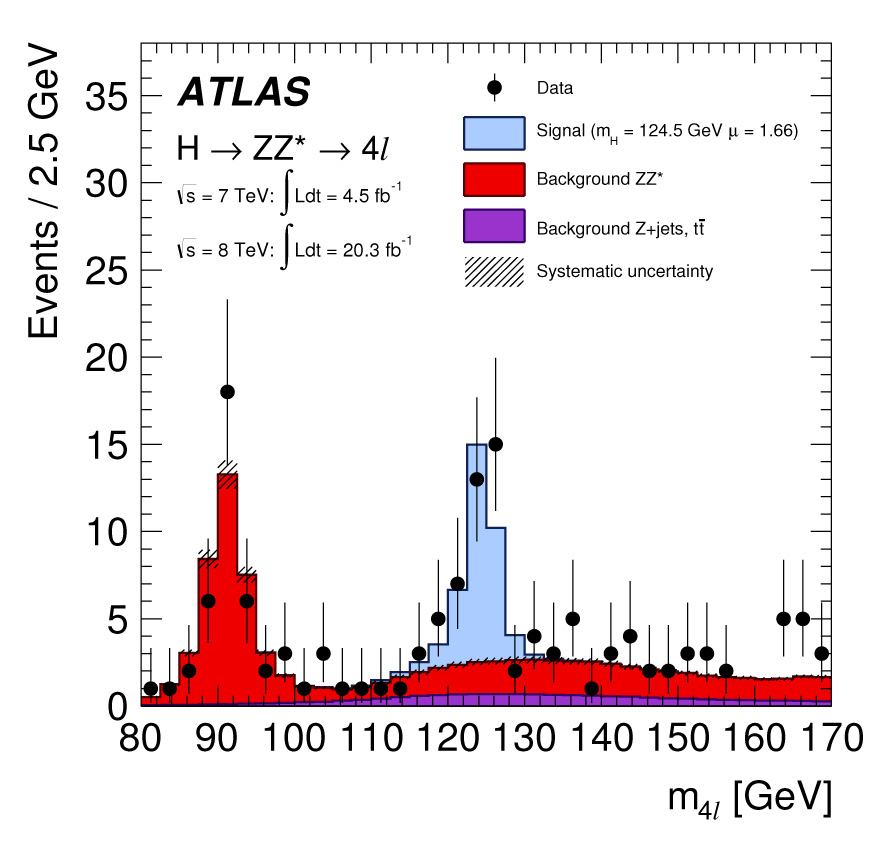}
\vspace{-2mm}
\caption{An invariant mass ($m_{4l}$) distribution in decay $H\rightarrow ZZ^{*}\rightarrow 4 l$ for the 
combined 
$\sqrt{s} = 7$ TeV and $\sqrt{s} = 8$ TeV data and the mass range 80 -- 170 GeV \cite{Aad:2014aba,Aad:2014eva}.}
\label{m4l}
\end{center}
\end{figure}

In every measurement, a key problem is to distinguish and separate the true signal events looked for or expected in a given process from the fake, or the background, ones coming 
from other processes or reasons. To solve that it is necessary to estimate the expected background composition and yield in the process measured. This estimation can be done via a Monte Carlo simulation normalized to the SM theoretical predictions (usually for electroweak-related processes) or by using data (usually for QCD-related processes). Backgrounds are classified into irreducible or reducible ones (see, for example, \cite{Aad:2012tfa,van:2007ictp}, for more details). The backgrounds of the first type are those events containing the same final states as those of the signals, while the backgrounds of the second type are  those events with final states mistreated as the true ones of the signals. Because, as mentioned earlier, the decays $H\rightarrow \gamma \gamma$ and $H\rightarrow ZZ \rightarrow 4l$ with their advantages are ``golden" channels for hunting the Higgs boson \cite{Aad:2012tfa,Chatrchyan:2012ufa}, here we briefly discuss the backgrounds (see Fig. \ref{mgamma} and Fig. \ref{m4l} for a quantitative imagination) in these two Higgs decay channels (see, for example, \cite{van:2007ictp}, for other channels). \\

For the channel $H\rightarrow \gamma \gamma$, the irreducible backgrounds consist of the genuine photon pairs produced in Born- ($qq\rightarrow \gamma\gamma$), box- ($gg\rightarrow \gamma\gamma$) and quark bremsstrahlung ($qg\rightarrow q\gamma\rightarrow \gamma\gamma$, $gg\rightarrow jj\gamma\gamma$) processes, where $j$ denotes a jet, while the reducible backgrounds consist of $\gamma$-jet- and jet-jet events in which one or two jets are misidentified as photons, or electrons in the decay $Z\rightarrow ee$ misidentified as photons. For the channel $H\rightarrow ZZ \rightarrow 4l$, the irreducible backgrounds come from $ZZ^*$ and $Z\gamma^*$ continuum productions including those in which one of the $Z$ decays into a pair of $\tau$ leptons which subsequently decay into lighter leptons. The reducible backgrounds for this channel consist of $4l$ produced from $tt$ and $Z+\mbox{jets}$ (the latter, for the final states $ll+\mu\mu$, are mainly $Zbb$). More information of the backgrounds for these channels and processing in the ATLAS- and CMS experiments can be found in \cite{Aad:2012tfa,Chatrchyan:2012ufa}.\\

\begin{figure}[htbp]
\begin{center}
\includegraphics[width=.6\textwidth, angle = 0]{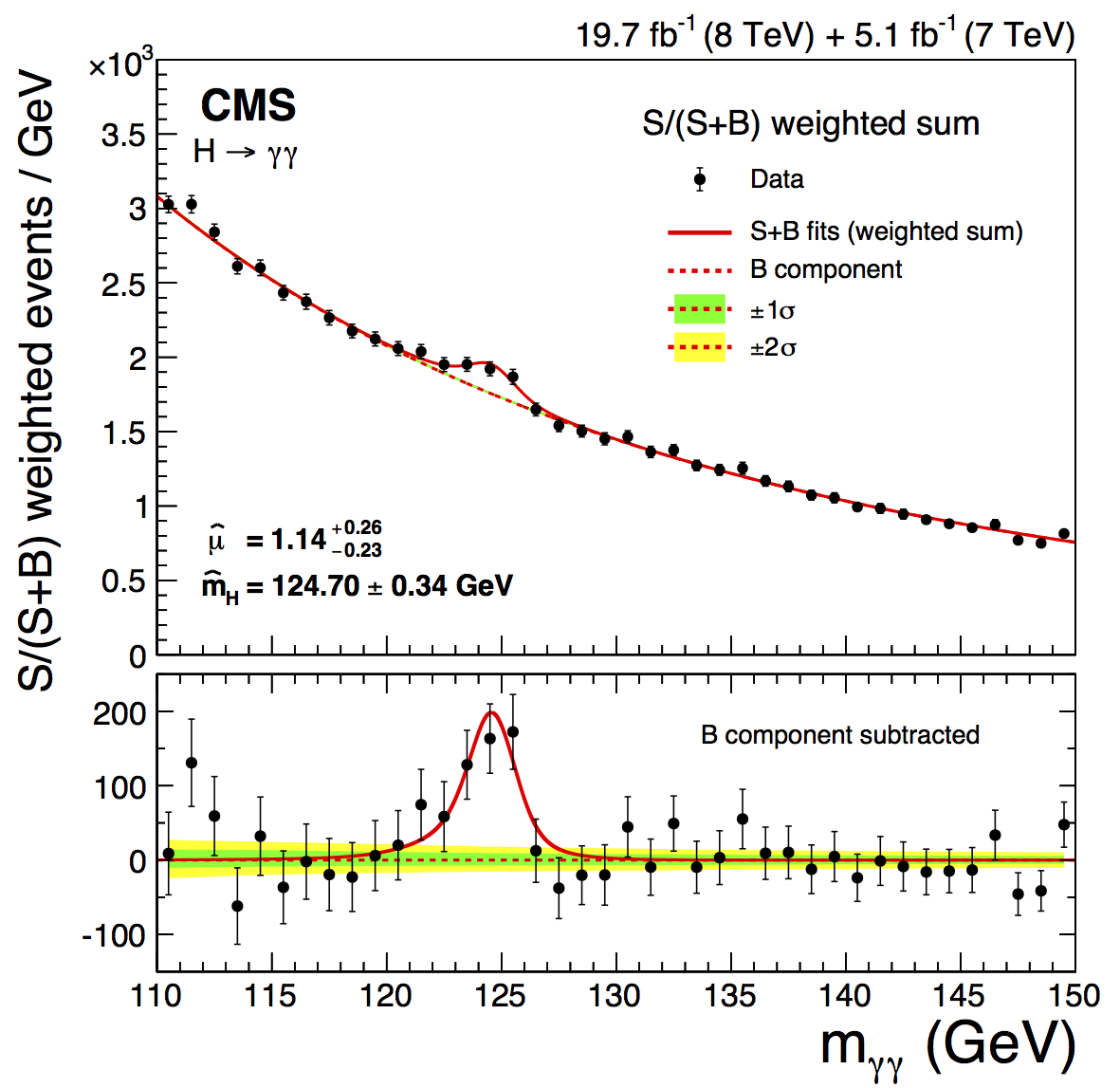}
\vspace{-2mm}
\caption{A combined diphoton Higgs mass spectrum measured by CMS, where the lower panel shows a distribution after a background subtraction \cite{Khachatryan:2014ira}.} 
\label{cms-mH}
\end{center}
\end{figure}

The Higgs mass found by ATLAS using the data collected in 2011 and 2012 from proton-proton collisions at the center-of-mass energy $\sqrt{s}= 7$ 
TeV and $\sqrt{s}= 8$ TeV with the integrated luminosity 4.5 fb$^{-1}$ and 20.3 fb$^{-1}$, has the following values \cite{Aad:2014aba}: 
\begin{equation}
m_H=125.98\pm 0.42(\mbox{stat.})\pm 0.28(\mbox{sys.})~ \mbox{GeV} =  125.98\pm 0.50 ~ \mbox{GeV},
\end{equation}
for the channel $H\rightarrow \gamma \gamma$ (see Fig. \ref{mgamma}) and 
\begin{equation}
m_H=124.51\pm 0.52(\mbox{stat.})\pm 0.06(\mbox{sys.})~ \mbox{GeV}= 124.51\pm 0.52 ~ \mbox{GeV},
\end{equation}
for the channel $H\rightarrow ZZ \rightarrow 4l$ (see Fig. \ref{m4l}). The difference between these mass measurements has a significance of about 2$\sigma$ corresponding to a probability of about 4.8$\%$. The combined result \cite{Aad:2014aba} obtained at 
$\sqrt{s}= 7$ TeV and $\sqrt{s}= 8$ TeV with the integrated luminosity 25 fb$^{-1}$ is 
\begin{equation}
m_H=125.36\pm 0.37(\mbox{stat.})\pm 0.18(\mbox{sys.})~ \mbox{GeV} = 125.36\pm 0.41 ~ \mbox{GeV}. 
\end{equation}
The corresponding results from the CMS read \cite{cms-mH,Khachatryan:2014ira}
\begin{equation}
m_H=124.70\pm 0.31(\mbox{stat.})\pm 0.15(\mbox{sys.})~ \mbox{GeV} =  124.70^{+0.35}_{-0.34} ~ \mbox{GeV},
\end{equation}
for the channel $H\rightarrow \gamma \gamma$,  
\begin{equation} 
m_H=125.6\pm 0.4(\mbox{stat.})\pm 0.2(\mbox{sys.})~ \mbox{GeV}= 125.6\pm 0.4 ~ \mbox{GeV},
\end{equation}
for the channel $H\rightarrow ZZ \rightarrow 4l$ and 
\begin{equation}
m_H=125.03^{+0.26+0.13}_{-0.27-0.15}~ \mbox{GeV} = 125.03\pm 0.30 ~ \mbox{GeV}. 
\end{equation}
for the combined  mass. As an illustration, a combined Higgs mass spectrum measured by CMS is depicted in Fig. \ref{cms-mH}.  
 A naive combination of the two Higgs masses obtained by ATLAS and CMS gives \cite{Ellis:2014cga}
\begin{equation}
m_H=125.15\pm 0.24~ \mbox{GeV}. 
\end{equation}

New measurements reported recently by ATLAS \cite{HWW:2014} showed that the Higgs boson was observed in the 
decay $H\rightarrow WW$ at a level of significance 6.1$\sigma$ (compared with the corresponding SM expected value 
5.8$\sigma$) and the recent observation by CMS for the channel $H\rightarrow \gamma \gamma$ has reached the significance of 5.7$\sigma$ (compared with the corresponding SM expected value 5.2$\sigma$)\cite{Khachatryan:2014ira}. More precise measurements and beautiful results are anticipated during the Run 2 of the LHC \cite{Ellis:2014cga}.\\

It is very important to look at the signal strengths of the observed channels. A signal strength by definition is a ratio 
$\mu=\sigma/ \sigma_{SM}$ between a measured value $\sigma$ and an SM theoretical value $\sigma_{SM}$ of a cross section. Therefore, a signal strength would equal 1 in an ideal case if the SM is a perfectly correct model. 
The signal strengths measured by ATLAS for the $\gamma\gamma$-, $ZZ$- and $WW$ channels of the Higgs decays 
are $\mu = 1.17\pm 0.27$ (for $H\rightarrow \gamma\gamma$) 
\cite{Aad:2014eha}, $\mu=1.44^{+0.40}_{-0.33}$ 
(for $H\rightarrow ZZ$) \cite{Aad:2014eva} and $\mu=1.08^{+0.22}_{-0.20}$ (for $H\rightarrow WW)$ 
\cite{HWW:2014} which are in good agreement with the SM.\\

The above-mentioned channels show that the new boson is coupled to the massive gauge bosons (directly) 
and photons (indirectly) as predicted by the SM. To see if it is the Higgs boson the next step which must be done 
is to check if it is also coupled to fermions, as predicted by the SM. As the coupling $Htt$ of the 
Higgs boson to the top quarks can be studied via the top fusion 
mentioned above, it remains to study the couplings of the Higgs boson to other, specially, down-type, 
fermions. ATLAS has made this study on the potential decay channels  
$H\rightarrow \tau\tau$ and $H\rightarrow bb$ for $\sqrt{s}= 7$ TeV (4.5 fb$^{-1}$ and 4.7 fb$^{-1}$, resp.) 
and $\sqrt{s}= 8$ TeV (20.3 fb$^{-1}$) \cite{Nagai:2013xwa, Htautau:2014}. 
 A similar study \cite{Chatrchyan:2014nva,Chatrchyan:2014vua,CMS:2014ega} has been made by CMS at the same energies but with 
other integrated luminosities which are 4.9 fb$^{-1}$ ($\sqrt{s}= 7$ TeV) and 19.7 fb$^{-1}$ 
(at $\sqrt{s}= 8$ TeV) for $H\rightarrow \tau\tau$; and 5.1 fb$^{-1}$ (at $\sqrt{s}= 7$ TeV) and 
18.9 fb$^{-1}$ (at $\sqrt{s}= 7$ TeV) for $H\rightarrow bb$.
Other decay channels of the Higgs boson into fermions are either 
low ranked because of small BR's (e.g., $H\rightarrow \mu\mu$) or kinematically impossible ($H\rightarrow tt$). Let us now consider the channels $H\rightarrow bb$ and $H\rightarrow \tau\tau$.\\

For the channel $H\rightarrow bb$ with $H$ produced in association with $W/Z$ (\textit{Higgs strahlung}) and $m_H=125.36$ GeV, 
according to ATLAS the observed significance of an event excess over the background is only 1.4$\sigma$ 
(compared with the expected 2.6$\sigma$) and the signal strength is $\mu = 0.52\pm 0.32 (\mbox{stat.}) \pm 0.24(\mbox{syst.})$ is quite small \cite{Aad:2014xzb} but the corresponding result of CMS is a bit better, i.e., the signal significance and strength are 2.2$\sigma$ and $\mu = 1.3^{+0.7}_{-0.6}$, respectively \cite{Nagai:2013xwa}. Both ATLAS and CMS have also studied the channel $H\rightarrow bb$ with $H$ produced from top fusions but neither of them has found a significant signal \cite{Nagai:2013xwa}. The ATLAS result for the channel $H\rightarrow \tau\tau$ is more convincing than that for $H\rightarrow bb$ as the observed deviation from the background is 
4.5$\sigma$ (compared with the expected 3.5$\sigma$) and the signal strength is $\mu = 1.42
^{+0:44}_{-0.38}$ \cite{Htautau:2014}. Although the signal strengths for the last two channels have 
significant deviations compared to those for other channels, the combined signal strength for all channels 
$H\rightarrow \gamma\gamma, ZZ, WW, bb$ and $\tau\tau$ measured 
by ATLAS, 
\begin{equation}
\mu=1.30\pm 0.12\pm 0.10\pm 0.009,
\end{equation}
is still in a quite good agreement with the corresponding values from CMS,
\begin{equation}
\mu=1.00\pm 0.09^{+0.008}_{-0.07}\pm 0.07,
\end{equation} 
Tevatron 
and the SM \cite{Ellis:2014cga}. \\

\begin{figure}
\begin{center}
 \includegraphics[width=.32\textwidth, angle = 0]{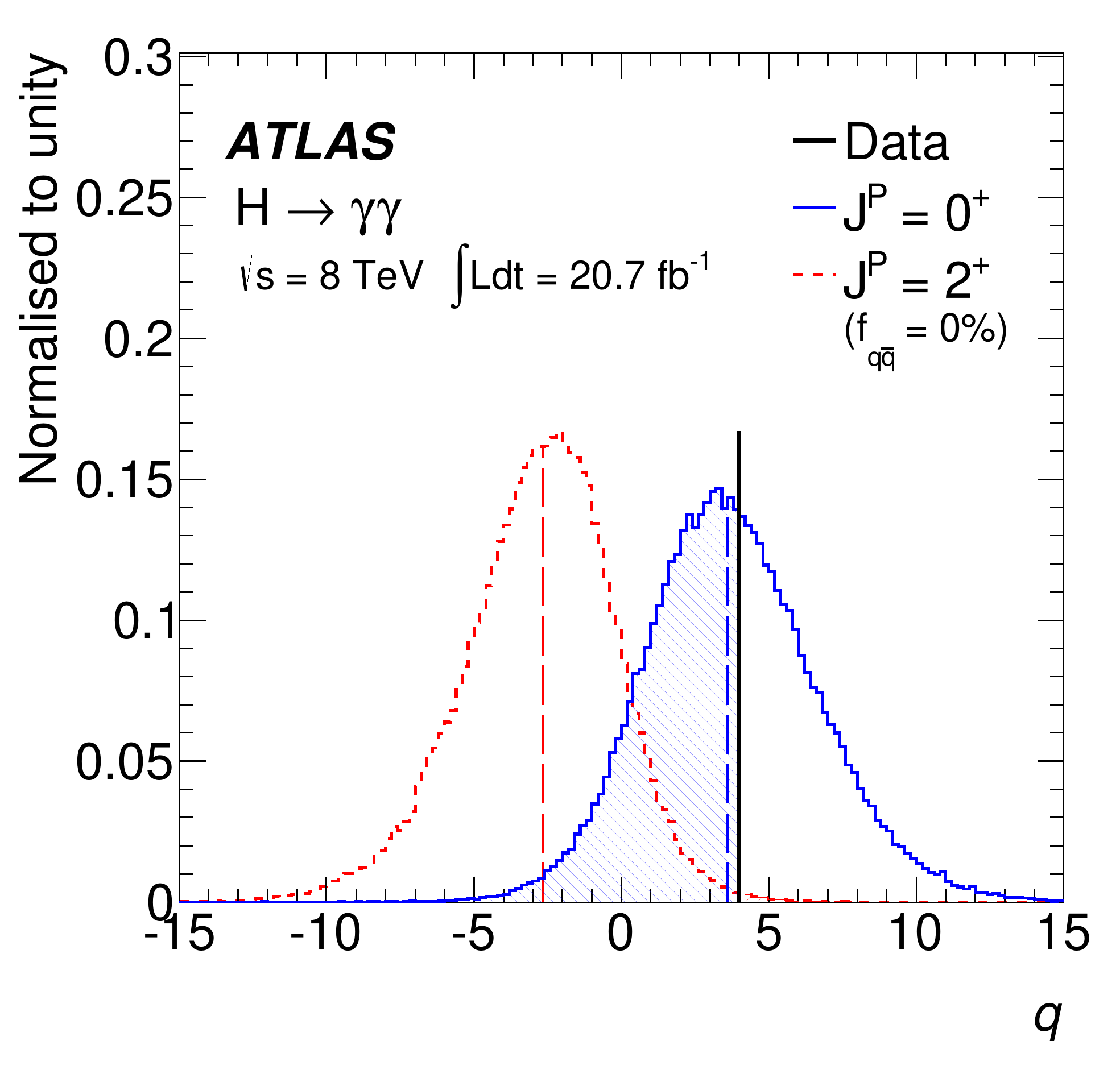} ~ \includegraphics[width=.32\textwidth, 
angle = 0]{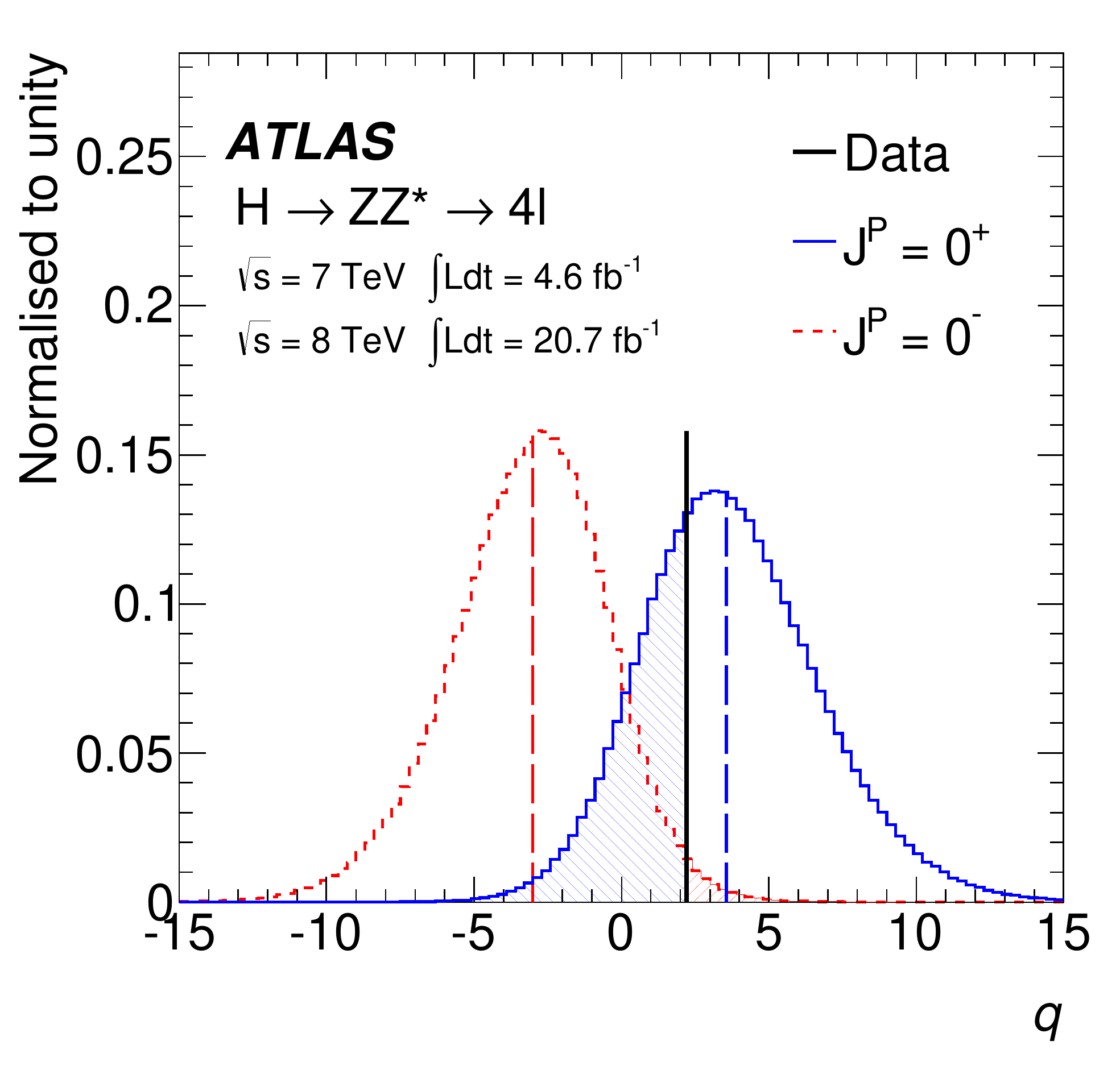}~ \includegraphics[width=.32\textwidth, angle = 0]{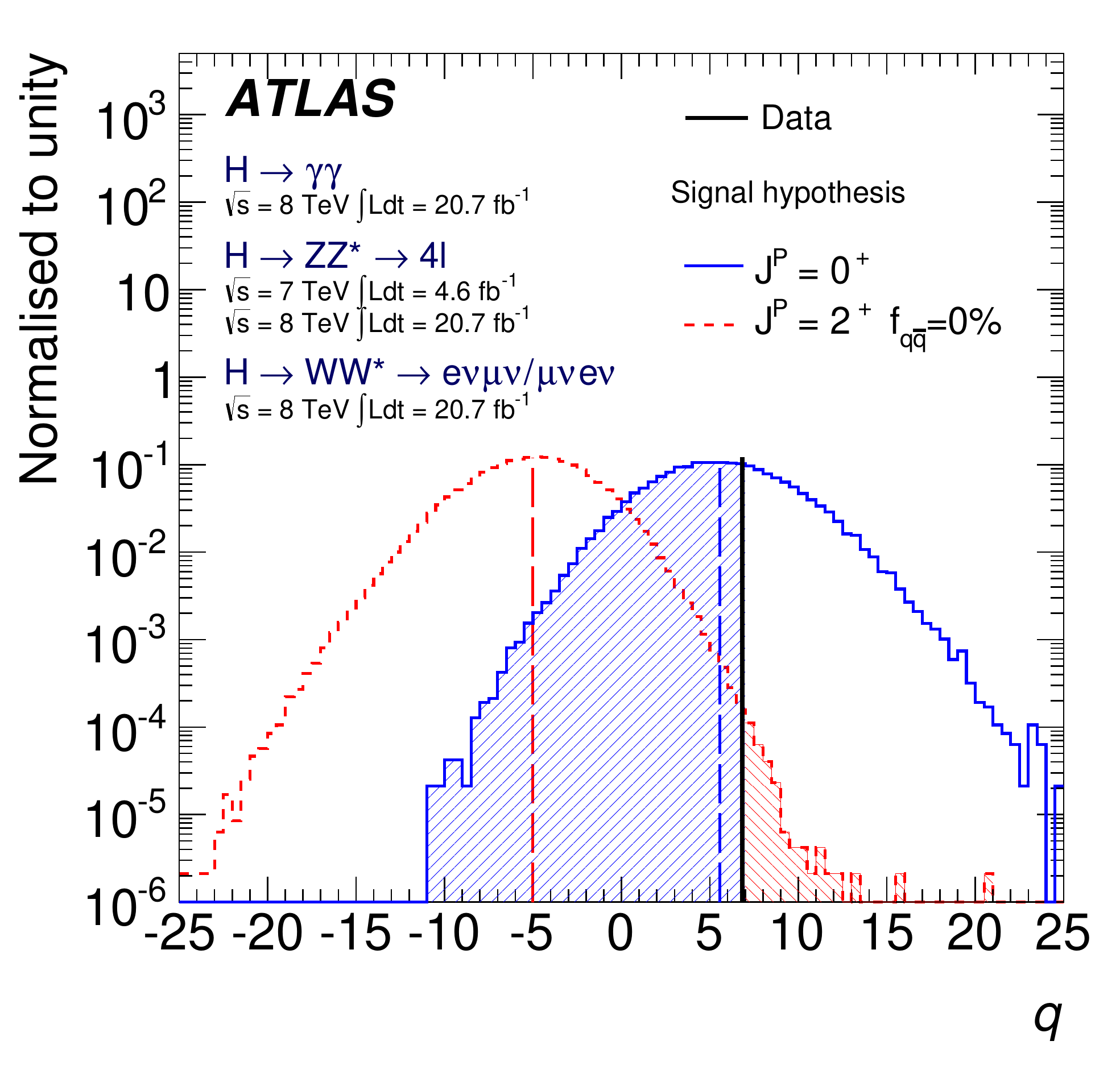}
 \caption{Examples for expected distributions of the logarithm of the ratio of profiled 
likelihoods ($q$), under the $J^P = 0^+$ hypothesis in comparison with other spin-parity hypotheses (the observed
values are shown by the vertical solid lines) \cite{ATLAS:2013mla,Aad:2013xqa}.}
 \label{Hspin}
 \end{center}
\end{figure}

In order to conclude whether the new particle is really or not the long-sought Higgs boson one more step 
which must be also done is to determine its spin and parity (spin-parity, $J^P$). First of all, as this particle 
decays into a pair of gauge bosons it can be neither a fermion nor a spin-1 particle. Therefore, it 
remains to check if it has a spin-0 or spin-2. This problem has been investigated by ATLAS (and also by 
CMS) via the decay channels $H\rightarrow \gamma \gamma$, $H\rightarrow ZZ \rightarrow 4l$ and 
$H\rightarrow WW \rightarrow l\nu l\nu$ at $\sqrt{s}= 8$ TeV (20.7 fb$^{-1}$) and the channel 
$H\rightarrow ZZ \rightarrow 4l$ at $\sqrt{s}= 7$ TeV (4.6 fb$^{-1}$). These investigations 
(see, for example, Fig. \ref{Hspin}) have given a strong evidence for the scalar (spin-0 and 
positive-parity) nature of the newly discovered particle as expected for the Higgs boson \cite{ATLAS:2013mla,Aad:2013xqa}.
%
%

\section*{Conclusions}
~

We have briefly reviewed some general lines and progress in the  history of the Higgs boson 
introduction and search as well as 
the main achievements in studying the new boson discovered in 2012 by the LHC collaborations 
ATLAS and CMS. The investigation by several collaborations (ATLAS, CMS and Tevatron) has shown 
with high confidence that the newly discovered boson is a scalar ($J^P=0^+$) having a mass about 
125 GeV and its couplings to other bosons and fermions checked in high precision are compatible 
with the standard model Higgs boson. These couplings are very important as they give rise to 
particle masses. Here, for illustrations, most of results are taken from ATLAS but they could be also taken from CMS which has similar missions and results with ATLAS. \\

The current results have been obtained by analyzing two and a half times more data than that available by 04 July 2012 when the discovery was announced. The observation has become more and more  convincing, for example, the confidence of the observation (by ATLAS) of the Higgs boson decaying into two photons and into two Z bosons has risen to 10 $\sigma$ \cite{atlas}. We hope to see more precise measurements and, maybe, new exciting results, in 
the coming time after the LHC starts its Run 2 in 2015 when the LHC is expected to reach a collision energy of 13 TeV which is a new record (the design collision energy 14 TeV could be reached sometime later). However, the results obtained so far by the collaborations ATLAS and CMS are (almost) enough to confirm that the newly discovered boson is exactly the 50-year long sought Higgs boson of the standard model (despite some doubt raised recently that the ``new boson" could  be treated as a resonance which is an iso-singlet scalar in a technicolor model \cite{Belyaev:2013ida}). \\

In the framework of this review, because of lack of space and because our purpose here is to 
see if the new boson discovered by ATLAS and CMS is the SM Higgs boson, we do not discuss physics beyond the SM  \footnote{For search for new physics by ATLAS and CMS, see, for example, recent papers \cite{Aad:2014yja,CMS:2015ula}.} which may need no Higgs field (in the so-called Higgs-less models) or more than one Higgs fields with different structures, i.e., Higgs bosons with different masses and properties.\\

The discovery of the Higgs boson, along with other achievements of the LHC program, is a triumph 
of not only the standard model and BEH mechanism but also science and technology in general and 
represents a nice fruit of international cooperation, which is always essential
in high-energy physics.

\section*{Acknowledgement} 
This work is supported by Vietnam's National Foundation for Science and Technology Development 
(NAFOSTED) under the grant No 103.03-2012.49. The authors would like to thank Patrick Aurenche for reading the manuscript and useful comments. 

%
%

\end{document}